\documentclass[twocolumn,showpacs,preprintnumbers,amsmath,amssymb,prl]{revtex4}
\usepackage{graphicx}
\usepackage{dcolumn}
\usepackage{bm}

\begin{document}

\title{
Signatures of a Noise-Induced Quantum Phase Transition \\
in a Mesoscopic Metal Ring }

\author{Ning-Hua Tong and Matthias Vojta} \affiliation{ Institut f\"{u}r Theorie der Kondensierten
 Materie, Universit\"{a}t Karlsruhe, 76128 Karlsruhe, Germany }

\date{\today}

\begin{abstract}
We study a mesoscopic ring with an in-line quantum dot threaded by an Aharonov-Bohm flux.
Zero-point fluctuations of the electromagnetic environment
capacitively coupled to the ring,
with $\omega^s$ spectral density,
can suppress tunneling through the dot,
resulting in a quantum phase transition from an unpolarized to a
polarized phase.
We show that robust signatures of such a transition can be found
in the response of the persistent current in the ring
to the external flux as well as to the bias between the dot and the arm.
Particular attention is paid to the experimentally relevant cases of ohmic ($s=1$)
and subohmic ($s=1/2$) noise.
\end{abstract}

\maketitle

The persistent current in a mesoscopic ring,
penetrated by an Aharonov-Bohm flux, has been studied intensively in
the past twenty years~\cite{Buettiker1}. This equilibrium current is
an indication of quantum coherent motion of charge carriers in the
ring. Recent studies show that the zero-point electromagnetic
fluctuations in the leads can suppress quantum coherence, and
effectively decrease the magnitude of the persistent
current~\cite{Cedraschi1}.

Theoretically, it is known that environmental dissipation can even
cause {\em qualitative} changes in the ground-state properties,
i.e., it can drive the system across a quantum phase transition
(QPT) into a dissipation-dominated phase. A popular model system is
the spin-boson model \cite{Leggett1,Weiss1}, describing a two-level
``impurity'' linearly coupled to a bath of harmonic oscillators.
Here, a QPT between a delocalized and a localized phase is found
both for ohmic and subohmic damping \cite{Vojta1}.
The more complicated Bose-Fermi Kondo model, with an additional
fermionic bath, also shows QPT between different
ground states \cite{Lehur1,Kirchner1,Glossop1,Vojta2}. Several
experiments have been suggested to observe such environment-induced
QPT, utilizing single-electron transistors with electromagnetic
noise \cite{Lehur1,Furusaki1,Lehur2}, quantum dots coupled to
ferromagnetic leads \cite{Kirchner1}, or cold atoms in optical
lattices \cite{Recati1}.

In this paper, we propose to use the persistent current in a
metallic ring as a detector of a noise-induced QPT.
A suitable setup \cite{Cedraschi1}
consists of a ring with a small in-line quantum dot, capacitively
coupled to an external circuit with dissipative impedance, Fig. 1.
The key idea is that tunneling through the barriers can be
effectively suppressed by charge fluctuations coupled to the ring
electrons. Whereas Ref.~\onlinecite{Cedraschi1} studied the
magnitude and the fluctuations of the persistent current in the
presence of ohmic noise, we focus here on the QPT caused by the
dissipative environment. Modeling the external circuit by a $RLC$
transmission line, we put emphasis on the cases of ohmic dissipation
and a subohmic dissipation with $\sqrt{\omega}$ spectrum density.
They can be realized as $LC$-dominant and $R$-dominant limit of the
transmission line, respectively \cite{Devoret1}. In both cases we
find robust signatures of a QPT between an unpolarized phase at
small dissipation and a polarized phase at large dissipation -- in
the latter, the charge state on the dot is doubly degenerate.
Unlike previous proposals for detecting noise-induced QPT, our setup
can be realized with either ohmic or subohmic dissipation, and allows to
assess the influence of finite temperature and bias.
\begin{figure}[b!]
\begin{center}
\includegraphics[width=1.6in, height=1.3in]{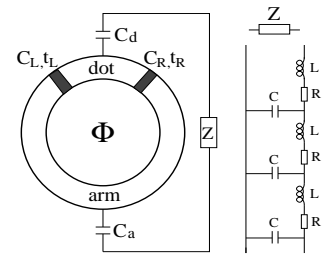}
\vspace*{-5pt}
\end{center}
\caption{
\label{circuit2}
Ring with an in-line dot subject to a flux $\Phi$ and capacitively
coupled to an external impedance $Z$. The dot is coupled to the left and
right part of the arm through barriers with tunneling strength $t_L$ and $t_R$,
and effective capacitors $C_L$ and $C_R$.
The impedance $Z$ is modeled by an infinite $RLC$ transmission line.}
\end{figure}

\emph{Model.} The setup, shown in Fig. 1, consists of a polarizable
environment coupled to the ring--dot system through the capacitors
$C_d$ and $C_a$. For simplicity we assume spin-polarized electrons;
adding the electronic spin degree of freedom would enhance the
magnitude of persistent current \cite{Buettiker2}, but not
qualitatively change the physics studied here. Let us first consider
the Hamiltonian for charges on the ring and the dot. In a small dot
the level spacing, $\Delta_d$, will be larger than the tunneling
amplitudes $t_L$ and $t_R$. Regarding the level spacing in the arm,
$\Delta_a$, we will separately analyze the cases of a small ring,
$\Delta_a \gg t_L, t_R$, and a large ring, $\Delta_a \ll t_L, t_R$.
For a small ring, low-temperature charge transport occurs only
between the two topmost energy levels of the dot and the arm
\cite{Cedraschi1}. This leads to a two-level system $H_{\rm tun}=
\frac{\epsilon}{2} \sigma_z - \frac{\Delta}{2} \sigma_{x}$, with
$\epsilon$ being the chemical potential difference between the dot
and the arm, and $\sigma_{x,z}$ are Pauli matrices. The tunneling
strength is $\Delta= 2\sqrt{ t_{L}^2+t_{R}^2 + 2 \lambda t_L t_R
\cos \varphi }$, where $\varphi=2\pi \Phi/\Phi_0$ is the external
flux in unit of the flux quantum $\Phi_0=hc/e$, and $\lambda= 1$
$(-1)$ for an odd (even) number of electrons on the ring. $\sigma_z$
is related to the charge on the dot by $Q_d= \sigma_z e/2  +
\left(N+1/2 \right)e$, assuming $N$ permanently occupied states. In
the opposite case of a large ring, the energy spectrum in the arm is
continuous, and progress can be made via bosonization of the arm
electrons. This leads to a coupling between the pseudospin
$\sigma_z$ and the charge density fluctuations in the arm, $H_{\rm
ohm}=v_{F} \sum_{k>0} k b_{k}^{\dagger}b_{k} -v_{F} \sigma_z
\sum_{k>0} \sqrt{\pi k /2L} (b_{k}+b_{k}^{\dagger}) $
\cite{Furusaki1}. It describes an ohmic bath $J_{\rm
ch}(\omega)=2\pi \alpha_{\rm ch} \omega\Theta(\omega_c-\omega)$ with
$\alpha_{\rm ch}=1/2$. The cutoff is $\omega_c = v_{F}k_{c}$, and
$v_{F}$ the Fermi velocity. In the tunneling part, $H_{\rm tun}$, a
factor $\sqrt{\rho \omega_{c}/2\pi}$ now appears in the expression
of $\Delta$, where $\rho$ is the density of states at Fermi energy
in the arm. Thus, in both the small-ring and large-ring cases we
arrive at a two-level system $H_{\rm tun}$, but supplemented with an
additional ohmic bath in the large-ring case.

Now we consider the electromagnetic environment. The $RLC$
transmission line, connecting to the dot and the arm, is described
by the distributed resistances $R$, inductances $L$, and
capacitances $C$ per unit length (Fig.1). A Hamiltonian description
of the environment can be obtained for an effective $LC$
transmission line ($R=0$). Allowing for an arbitrary distribution of
$L_n$, $C_n$ ($n=1,2, \ldots$), the Hamiltonian reads
\cite{Cedraschi1} $H_{\rm em}=Q_0^2/(2 C_0) +\sum_{n=1}^{\infty}
\left[ Q_n^{2}/(2 C_n)  + \left(\phi_n - \phi_{n-1} \right)^2/(2
L_n)  \right] $. The impedance is $Z\left(\omega \right) =i\omega
L_1 + (i \omega C_1+ i \omega L_2 +   ( i \omega C_2 + ...
)^{-1})^{-1})^{-1}$. $Q_0$ in $H_{\rm em}$ is the charge on the
capacitors $C_a$ and $C_d$, and $C_0^{-1}= C_a^{-1} + C_d^{-1} +(C_L
+C_R )^{-1}$. The operators $\phi_n$ and $Q_n$ ($n \ge 0$) are
conjugate operators satisfying $[ \phi_n, Q_m] = \delta_{nm}$. The
system--environment coupling originates from the Coulomb
interactions, i.e., the charging energy of the capacitors $C_L$ and
$C_R$, $H_{\rm coup}=e/[2\left( C_L + C_R \right)]Q_0 \sigma_z$.
Hence the full Hamiltonian for the small-ring case reads $H=H_{\rm
tun}+H_{\rm coup}+H_{\rm em}$, and for the large-ring case $H=H_{\rm
tun}+H_{\rm coup}+H_{\rm em}+H_{\rm ohm}$.

To study the ground-state properties of $H$, we use Wilson's
numerical renormalization group (NRG) method \cite{Wilson1}.
NRG is a non-perturbative method for impurity problems and was recently extended to
impurity problems with bosonic bath \cite{Bulla1}.
We diagonalize the bath degrees of freedom in $H_{\rm em}$ and transform $H$
into the standard form of the spin-boson model
\begin{eqnarray}
   H_{sb} &=& \frac{\epsilon}{2}\sigma_z - \frac{\Delta}{2}\sigma_x
                 + \frac{\sigma_z}{2} \sum_{i} \lambda_{i} \left(a_{i}^{\dagger}+ a_{i} \right)
                 + \sum_{i} \omega_i a_{i}^{\dagger} a_{i}, \nonumber\\[-6pt]
\end{eqnarray}
with the spectrum $J( \omega )=\pi \sum_{i} \lambda_{i}^{2} \delta (\omega - \omega_i)$.
 For the small-ring case, $J(\omega)$ is solely from the
electromagnetic contribution $J_{\rm em}(\omega)$ evaluated below. For the large-ring case,
$J(\omega)=J_{\rm em}(\omega)+J_{\rm ch}(\omega)$
where $J_{\rm ch}(\omega)$ is from $H_{\rm ohm}$ as discussed above.

The electromagnetic contribution $J_{\rm e
m}(\omega)$ is related to the free retarded Green's function of $Q_0$
as $J_{\rm em} (\omega )=- e^2/ ( C_L+C_R )^2 Im \langle \langle Q_0 | Q_0 \rangle \rangle_{\omega+i \eta}$.
Since $\langle \langle Q_0 | Q_0 \rangle \rangle_{\omega}= -1/[i \omega Z (\omega) +C_0^{-1}]$,
we have
\begin{equation}
  J_{\rm em}\left( \omega \right)= - a \omega Re Z_t \left( \omega+ i \eta \right)    \, \, .
\end{equation}
Here $Z_t ( \omega )=1/ [Z^{-1} (\omega) + i \omega C_0 ]$ is the
total impedance describing the effective capacitor $C_0$ in parallel
with $Z ( \omega )$, and the coefficient $a=[ e C_0/(C_L + C_R)
]^{2}$. For the discrete transmission line in Fig. 1, $Z(\omega)$ is
a complicated function, which in the continuum limit simplifies to
$Z(\omega) = [(R + i\omega L)/i \omega C]^{1/2}$ \cite{Devoret1}. In
both cases, the asymptotic behavior of $J_{\rm em}$ is $J_{\rm em}=a
\sqrt{R/(2C)}\sqrt{\omega}$ for $\omega \ll R/L, 1/\sqrt{LC}$ and $
J_{\rm em}= a \sqrt{L/C}\omega$ for $1/\sqrt{LC} \gg \omega \gg
R/L$. The crossover energy scale $R/L$ separates the low-energy
subohmic $\sqrt{\omega}$ part from the high-energy ohmic $\omega$
part.

For large resistivity $R$ ($R$-dominant leads), the $\sqrt{\omega}$
part of the spectrum will dominate the dissipation. In the opposite
case ($LC$-dominant leads), $R/L$ is small, and the linear $\omega$
spectrum will be relevant in the experimentally accessible range of
energies and temperatures. For either type of the leads, the role of
the charge modes in the large-ring case, $J_{\rm ch}(\omega) \propto
\omega$, is only to renormalize the dissipation arising from $J_{\rm
em}(\omega)$ and will not change the quantum critical behavior. We
will therefore describe both small-ring and large-ring cases using a
``pure'' bath with a power-law spectral density $J( \omega )=2\pi
\alpha \omega^{s} \omega_c^{1-s} \Theta ( \omega_c- \omega)$, where
the exponent $s$ is determined by $J_{\rm em}(\omega)$. The
effective dissipation strength $\alpha$ is from $J_{\rm em}(\omega)$
in the small-ring case, and in the large-ring case will be further
renormalized by $J_{\rm ch}(\omega)$. $\omega_c$ is the effective
cutoff and taken as our energy unit.

The circulating persistent current operator is $I_c = (C_R I_L- C_L
I_R)/(C_{L}+C_{R})$, where $I_L$ and $I_R$ are operators for
particle current through the left and right tunnel barrier,
respectively \cite{Cedraschi2}.
Due to $\langle \sigma_y \rangle =0$ the average displacement current $\langle I_L+I_R\rangle$
through the external loop vanishes, and the expectation value of $I_c$ reads
$I=\langle I_L \rangle=-\langle I_R \rangle=I_0 \langle \sigma_x \rangle$.
$I_0=-e/2 \partial \Delta / \partial \varphi$ is the
persistent current in the dissipationless limit.

{\it Phase transition.}
For a given experimental configuration, the bath exponent $s$ and the
dissipation strength $\alpha$ are fixed.
$\Delta$ can be tuned in the range $[ |t_L -t_R|, t_L + t_R ]$ by scanning $\Phi$.
The bias $\epsilon$ is tunable by a gate voltage on the dot.
For both the ohmic and subohmic spin-boson models, a quantum critical point
$\Delta_c(\alpha)$ separates the localized ($\Delta < \Delta_c$) and delocalized
($\Delta > \Delta_c$) ground states \cite{Bulla1}.
In terms of the charges on the dot, these are the ``polarized'' and
``unpolarized'' states, respectively.
Note that $\langle \sigma_z\rangle$ can be viewed as an order parameter
for this transition, with $\langle \sigma_z\rangle\neq 0$ in the polarized
phase.
Whereas in the subohmic case a $\Delta_c$ exists for any $\alpha$
(with $\Delta_c\to 0$ as $\alpha\to 0$), in the ohmic case $\alpha >1$
is required for the existence of a localized phase.
Experimentally, one can cross the critical point by scanning $\Phi$,
provided that $|t_L-t_R| < \Delta_c < t_L+t_R$.

\emph{Results.}
Our results were obtained via the bosonic NRG for the spin-boson model (2),
using NRG parameters
$\Lambda=2$ (logarithmic discretization),
$M=80$ (kept states), and $N_b=6$ (local boson states) \cite{Bulla1}.
First we focus on the ground-state persistent current $I=\langle I_c \rangle$
and its response to $\Delta$.
The $\Delta$-response is related to
$\partial I(\Phi)/\partial \Phi$ which can be directly observed in
experiment \cite{Deblock1}.
With the periodic background $I_0$ removed, $I/I_0=\langle \sigma_x \rangle$ increases
monotonously as $\Delta$ increases and saturates in the
large-$\Delta$ limit.
(Note that $\langle \sigma_x \rangle\neq 0$ in both phases of the spin-boson model.)
For the subohmic bath, $0<s<1$, $I/I_0 \propto c(\alpha) \Delta$ in
the small-$\Delta$ limit, as shown in Fig. 2(a).
 $c(\alpha)$ increases with decreasing $\alpha$ and diverges
in the limit  $\alpha \rightarrow 0$.
For the ohmic case, $s=1$, our data extrapolated to $\Lambda=1$
are consistent with the perturbative results
$I/I_0 \propto \Delta^{\kappa (\alpha)}$
where  $\kappa(\alpha)=\alpha / (1-\alpha)$ for $\alpha < 1/2$ and
$\kappa(\alpha)=1$ for $\alpha \ge 1/2$ \cite{Cedraschi1,Weiss1}.

\begin{figure}[t!]
\begin{center}
\includegraphics[width=3in, height=1.4in]{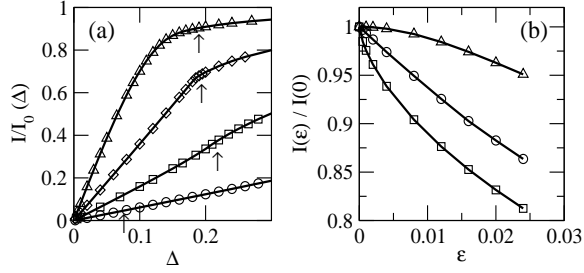}
\vspace*{-10pt}
\end{center}
\caption{(a) Rescaled persistent current as function of $\Delta$,
for $S=0.2, \alpha=0.03$; $S=0.5, \alpha=0.15$;
$S=0.8, \alpha=0.5$; and $S=1.0, \alpha=1.2$ (from top to bottom).
The arrows mark the positions of $\Delta_c$.
(b) Rescaled persistent current as function of bias
$\epsilon$, at $S=0.5$, $\alpha=0.15$ for $\Delta=0.3$
(triangles), $\Delta=\Delta_c=0.18463$ (squares),
and $\Delta=0.05$ (circles).
}
\vspace{-8pt}
\end{figure}

Approaching the quantum critical point $\Delta=\Delta_c$,
a singular contribution arises: $I/I_0=(I/I_0)_c + c_{+/-}
|\Delta-\Delta_c|^{\theta}+ (I/I_0)_{\rm reg}(\Delta-\Delta_c)$. The
exponent $\theta$ can be related to the longitudinal susceptibility
exponent $\gamma_{33}$ \cite{Vojta1,gamma1}, defined as $\partial
\langle \sigma_z \rangle/\partial \epsilon \propto
|\Delta-\Delta_c|^{-\gamma_{33}}$, using hyperscaling (see below),
with the result $\theta=\gamma_{33}/s-1$. Since $\theta>1$, except
for $s=1/2$, $I/I_0$ near $\Delta_c$ is always dominated by the
linear term of regular contribution $(I/I_0)_{\rm reg}$, masking the
power law in the raw data. However, the susceptibility $\chi_{11}
\equiv \partial \sigma_x/ \partial \Delta =\partial (I/I_0)/
\partial \Delta$ shows clear singularity in the range $ 1/3 < s <
0.92 $ where $\theta-1 <1$. For the $R$-dominant leads, $s=1/2$, a
kink appears in the $I/I_0$ curve at $\Delta_c$, manifesting the
exponent $\theta=1$ \cite{gamma1}, see Fig. 2(a). Correspondingly,
there is a finite jump at $\Delta_c$ in the $\chi_{11}(\Delta)$
curve.

The singular behavior at the QPT can be nicely extracted from
the persistent current as a function of the bias $\epsilon$.
$I(\epsilon)$ is a decreasing function and $I(0)$ depends on both
$\alpha$ and $\Delta$.
$I(\epsilon)$ has a distinct small-$\epsilon$ behavior in the different phases.
For $\Delta > \Delta_c$, $I(\epsilon)/I(0) \propto 1- c \epsilon^{2}$ with a
coefficient $c$ depending on $\Delta$ and $\alpha$;
in the decoupled limit $c=1/(2\Delta^2)$.
For $\Delta < \Delta_c$,
$I(\epsilon)$ has a finite non-universal slope at $\epsilon=0$.
At the critical point $\Delta=\Delta_c$, a power law appears:
$1- I(\epsilon)/I(0) \propto |\epsilon|^{1/\delta_{13}}$.
Such distinct behaviors are shown in Fig. 2(b) for $s=1/2$.
For other $s$ values, while the behavior for the delocalized and localized phases is
similar, the slope of the critical curve at $\epsilon=0$ may be
zero, a finite value, or divergent, depending on whether $\delta_{13}(s)$ is smaller than,
equal to, or larger than unity.
The transition point $\delta_{13}=1$ is at $s=1/3$ (Fig. 4(a)).
For $s=1/2$, the exponent $\delta_{13}=3/2$, leading to a diverging
slope of the critical curve, Fig. 2(b).
\begin{figure}[t!]
\begin {center}
\includegraphics[width=3.0in, height=1.4in]{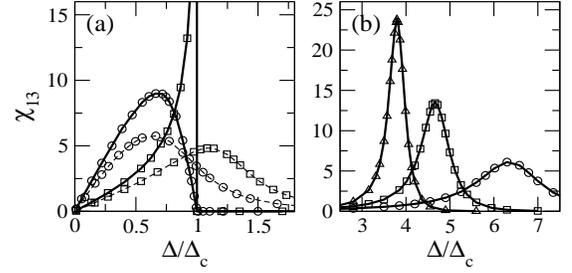}
\vspace*{-10pt}
\end{center}
\caption{
Susceptibility $\chi_{13}$ as function of the rescaled tunneling
 strength $\Delta/\Delta_c$. (a) subohmic case:
 $S=0.2$, $\alpha=0.03$ (circles) and
  $S=0.5$, $\alpha=0.15$ (squares). The solid lines are for $\epsilon=0$
   and dashed lines for $\epsilon=0.01$.
 (b) ohmic case:  $S=1$, $\alpha=1.2$. $\Delta_c \approx 0.0714$.
  For bias $\epsilon=10^{-4}$ (dots), $\epsilon=10^{-6}$
  (squares) and $\epsilon=10^{-8}$ (up triangles), respectively.
}
\vspace{-8pt}
\end{figure}

To characterize the response of the persistent current to the bias,
 we introduce a transverse susceptibility
  $\chi_{13} \equiv -\partial \langle \sigma_x \rangle / \partial \epsilon |_{\epsilon=0}
     = -(1/I_0) \partial I(\epsilon)/\partial \epsilon |_{\epsilon=0}$.
Not only can this quantity be measured from the linear response of the persistent current to
 the bias, it is also related to the flux induced capacitance
$C_{\Phi}=\partial \langle Q_d\rangle / \partial \Phi$ through
$C_{\Phi}=-e/(2\pi c)I_0 \chi_{13}$ \cite{Buettiker3}.
In the delocalized phase $\langle\sigma_z\rangle=0$,
meaning $\chi_{13}=C_{\Phi}=0$ for $\Delta > \Delta_c$.
$\chi_{13}$ is finite for $\Delta < \Delta_c$, and obeys a power law when
approaching the quantum critical point from this side,
$\chi_{13} \propto (\Delta_c-\Delta)^{-\gamma_{13}}$. $\gamma_{13}$ is the transverse
susceptibility exponent.
In Fig. 3(a), we show $\chi_{13}$ as a function of
$\Delta/\Delta_c$ for $s=0.2$ and $s=0.5$ (symbols with solid lines).
Since $\gamma_{13}$ changes sign at $s=1/3$, near the critical point
$\chi_{13}$ approaches zero for $s<1/3$ while it diverges for $s>1/3$.
This explains the different behavior of $s=0.2$ and $s=0.5$ curves in Fig. 3(a).

We examined the above scenario under a finite bias and finite temperatures.
As shown in Fig. 3(a) (symbols with dashed lines),
  a small bias $\epsilon=0.01$ smears the power-law singularity
 of $\chi_{13}$ and leads to a rounded peak structure in
  the $\chi_{13}(\Delta)$ curve.
  The deviation from the zero-bias curve is most prominent near $\Delta_c$.
  Turning on a finite temperature, $\chi_{13}$ are
  further suppressed on both sides of $\Delta_c$. However, in the regime
   $T < \epsilon$, the peak structure of $\chi_{13}$ curve
  is always a pronounced feature. This provides a signature of the
 quantum critical point which is robust against finite bias and finite temperature.

For the $LC$-dominant leads, $s=1$, a Kosterlitz-Thouless (K-T)
  transition occurs at a finite $\Delta_c$ only for $\alpha >1$. In this regime,
  $1-I(\epsilon)/I(0) \propto c\epsilon$ for both $\Delta < \Delta_c$ and
  $\Delta = \Delta_c$, consistent with $\delta_{13}(s\!=\!1)=1$ \cite{Bulla1}.
The transverse susceptibility follows
$\chi_{13} \propto \delta(\Delta-\Delta_c)$, instead of
the $(\Delta_c-\Delta)^{-1}$ divergence naively expected from
$\gamma_{13}\to 1$ as $s\to 1$.
  In Fig. 3(b), we plot the $\chi_{13}(\Delta)$ curve for $s=1$,
  $\alpha=1.2$ under different bias. The peak structure for finite
  bias evolves towards a $\delta$-peak in the small-$\epsilon$ limit.
  A finite bias shifts the peak position from $\Delta_c$ towards larger $\Delta$, and broadens
   the peak dramatically. These $\chi_{13}$ curves under finite bias thus
   present an well-defined precursor of the true QPT at zero bias.
For finite temperature, $T < \epsilon$, similar peaks are observed with
suppressed peak height.

\begin{figure}[t!]
\begin {center}
\includegraphics[width=3.0in, height=1.4in]{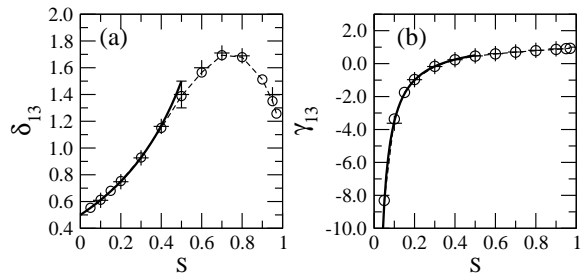}
\vspace*{-10pt}
\end{center}
\caption{
Critical exponents (a) $\delta_{13}$ and (b) $\gamma_{13}$ (crosses)
as function of $s$, obtained from NRG.
Also shown are the hyperscaling results
(a) $\gamma_{33}(1+s)/[2(\gamma_{33}-s)]$
and
(b)  $1-\gamma_{33}(1-s)/(2s)$
(circles with dashed line).
The solid lines are the values obtained from hyperscaling relation by setting
$\gamma_{33}=1$ in the $s < 1/2$ regime \cite{gamma1}.
}
\vspace{-8pt}
\end{figure}

Finally, in Fig. 4 we show the two critical exponents $\delta_{13}$ and $\gamma_{13}$.
A scaling ansatz for the free energy \cite{Vojta1} allows to derive
hyperscaling relations between critical exponents,
with the results $\delta_{13}=\gamma_{33}(1+s)/[2(\gamma_{33}-s)]$
and $\gamma_{13}=1-\gamma_{33}(1-s)/(2s)$, also shown in
Fig. 4 with $\gamma_{33}$ calculated by NRG.
The numerical consistency confirms the validity of hyperscaling
in the subohmic spin-boson model in the entire regime of $0 < s <1$ \cite{Vojta1}.
We also compare the numerical data with the hyperscaling results
obtained by setting $\gamma_{33}=1$ for $s\leq 1/2$.
Within numerical errors, most pronounced at $s=1/2$ due to
logarithmic corrections, our data support the analytical result $\gamma_{33}=1$ for
$0 < s < 1/2$ \cite{gamma1,Florens1}.
Note that at $s=1/3$, $\delta_{13}$ passes one and $\gamma_{13}$ passes zero,
leading to changes in the critical behavior of observables as discussed above.

Let us estimate the parameters in a typical experiment.
We take the surface plasma frequency $\omega_p = 10^{14} s^{-1}$ as the
cutoff energy and energy unit.
The upper boundary of the tunable dimensionless $\Delta$ is of the
order $10^{-3} - 10^{-2}$~\cite{Koenemann1}, and the lower
boundary can be very small for symmetric ring-dot system.
The dissipation strength $\alpha_{s=1} \sim 10^{-2}$ and $\alpha_{s=1/2}
\sim 10^{-3} - 10^{-2}$~\cite{Devoret1}.
For $s=1/2$ (R-dominant leads) this parameter space covers a sufficient region
of the phase diagram to make the observation of QPT feasible.
For $s=1$ (LC-dominant leads) the $\alpha$ is too small,
but the characteristic impedance $Z_0$ may be increased significantly
by optimizing the circuit design and selecting suitable lead materials.
Taking into account the $\alpha_{ch}=1/2$ dissipation from
charge fluctuations in the large-ring limit, it is still
possible to realize an ohmic dissipation source with $\alpha>1$, and
hence to observe the proposed features of the K-T transition.
For a micron-sized ring, an in-plane magnetic field of $1 - 10$ T will
effectively suppress the spin degrees of freedom to realize the spinless
electrons employed here.

\emph{Summary.}
We have identified robust signatures of a QPT, driven by
zero-temperature equilibrium environmental noise,
in a mesoscopic metal ring.
Suitable observables, showing quantum critical behavior, can be derived from
the persistent current through the ring.
We propose that this system can serve as a tool for studying the noise-induced QPT
and probing the low-energy electromagnetic fluctuations in a mesoscopic circuit.

The authors acknowledge helpful discussions with M. B\"{u}ttiker, R. Bulla,
C. H. Chung, S. Florens, R. Narayanan, G. Sch\"{o}n, and A. Shnirman.
This research was supported by the DFG through
the Graduiertenkolleg GRK 284 (NHT)
and the Center for Functional Nanostructures Karlsruhe (MV).


\begin{thebibliography}{99}
\bibitem{gamma1} Preliminary analytical results for the spin-boson model \protect\cite{Florens1}
show that $\gamma_{33}=1$ in the regime $0 < s  \leq 1/2$.
\end{thebibliography}
\end{document}